\documentclass[conference]{IEEEtran} 
   \usepackage{times}
   \usepackage{mathptm}
   \usepackage{caption}
   \usepackage{epsfig}
    \usepackage{subfigure}
   \usepackage{bbold}
   \usepackage{color}
   \usepackage{latexsym}
   \usepackage{pslatex}
   \usepackage{algorithmic}
   \usepackage{algorithm}
   \usepackage{graphicx}
  \usepackage{multirow}

\newtheorem{lemma}{Lemma}
\newtheorem{theorem}{Theorem}

\newcommand{\procname}[1]{\textsc{#1} }
\newcommand{\ALGORITHM}{\textbf{algorithm} }
\newcommand{\AND}{\textbf{and} }

\newcommand{\BL}{\textbf{begin\_loop} }

\newcommand{\ELSE}{\textbf{else} }

\newcommand{\END}{\textbf{end} }
\newcommand{\EL}{\textbf{end\_loop} }
\newcommand{\DO}{\textbf{do} }

\newcommand{\FOR}{\textbf{for} }

\newcommand{\IF}{\textbf{if} }

\newcommand{\OR}{\textbf{or} }

\newcommand{\RETURN}{\textbf{return} }

\newcommand{\THEN}{\textbf{then} }
\newcommand{\GO}{\textbf{go} }
\newcommand{\TO}{\textbf{to} }

\newcommand{\CONTINUE}{\textbf{continue} }

\newcommand{\true}{\mathsf{true}}
\newcommand{\false}{\mathsf{false}}

\begin{document}

\title{A Fast Heuristic Algorithm for Redundancy Removal}
\author{
\IEEEauthorblockN{Maxim Teslenko}
\IEEEauthorblockA{Ericsson Research\\ 
Ericsson AB\\
164 80 Stockholm, Sweden\\
maxim.teslenko@ericsson.com}
\and
\IEEEauthorblockN{Elena Dubrova}
\IEEEauthorblockA{Dept. of Electronic and Embedded Systems\\ 
Royal Institute of Technology - KTH\\
164 40 Stockholm, Sweden\\
dubrova@kth.se}
}

\maketitle

\begin{abstract}
Redundancy identification is an important step of the design flow that typically follows logic synthesis and optimization. In addition to reducing circuit area, power consumption, and delay, redundancy removal also improves testability.  All commercially available synthesis tools include a redundancy removal engine which is often run multiple times on the same netlist during optimization. This paper presents a fast heuristic algorithm for redundancy removal in combinational circuits. Our idea is to provide a quick partial solution which can be used for the intermediate redundancy removal runs instead of exact ATPG or SAT-based approaches. The presented approach has a higher implication power than the traditional heuristic algorithms, such as FIRE, e.g. on average it removes 37\% more redundancies than FIRE with no penalty in runtime. 
\end{abstract} 

\section{Introduction}
\label{into}

Combinational redundancy removal is an optimization problem that can
be formulated as follows: A gate or a net in a combinational circuit
is {\em redundant} if it can be removed without changing the functionality
of the circuit.
Very few, if any, synthesis tools
guarantee that the circuits they produce do not contain redundancy.
Unnecessary gates or connections are usually introduced by the
traditional optimization techniques such as factorization~\cite{BrM82}
or local transformations~\cite{darringer}.  In principle, it is
possible to restrict the transformations applied by a synthesis tool
to those that preserve non-redundancy of the original
circuit. However, it has been shown that redundancy gives an algorithm
a greater flexibility in restructuring the logic and more
possibilities to find a better implementation~\cite{SeD01}.

Since redundancy cannot always be avoided, 
all commercially available synthesis tools include a redundancy
removal engine that may be used multiple times on the same netlist
during optimization.  The presence of redundancy can cause several
problems. First, redundancy increases chip area, and may increase its
power consumption and propagation delay~\cite{LiLC95}. Second,
redundancy is the reason for undetectable faults in combinational
circuits.  Although undetectable faults do not affect the operation of
the circuit, they may block the detection of other faults and may
invalidate the completeness of a test set that was
generated~\cite{AbI92}.

In this paper, we consider two types of redundancy: (1)
redundancy associated with undetectable stuck-at faults, which do not
cause incorrect output values for any input assignment, and (2)
functional duplication, which occurs if different gates implement the
same function. {\em Automatic test patter generation (ATPG)} and {\em
fault-independent} methods target the first type of
redundancy. 

ATPG-based algorithms use exhaustive test pattern
generation to prove the undetectability of faults on redundant
lines~\cite{Fr67,ScA89,BeE02,BrD94}. They guarantee detection of all
such faults, but they have the exponential worst-case time complexity.

Fault-independent methods analyze the topology of a circuit without
targeting a specific fault. This can be done either by an explicit
analysis of reconvergent fanout regions, as in~\cite{HaM89}
and~\cite{MeA92}, or by propagating uncontrollability and
unobservability values, as in FIRE~\cite{IyA96} and its
extensions~\cite{GuH00,Hs02,ViSH05}. Although fault-independent
methods cannot determine all undetectable faults, they have an
advantage of the polynomial worst-case time complexity.

{\em Satisfiability checking (SAT)}~\cite{KiSSS97}, {\em Binary
Decision Diagram (BDD) sweeping}~\cite{kuehlmann} and {\em structural
hashing}~\cite{KuKr97} methods target the functional duplication type
of redundancy. SAT-based algorithms usually first partition all gates into
equivalence classes by random simulation, and then apply
satisfiability check for each pair in the class to verify equivalence.
BDD-sweeping algorithms build a binary decision diagram for
every gate in the circuit and merge gates with equivalent BDDs.  Both,
SAT and BDD-sweeping, guarantee detection of all functional
duplications, but they have the exponential worst-case time complexity.
Structural hashing can identify structurally isomorphic equivalent
vertices in the linear time.

This paper presents a redundancy identification and removal algorithm
which employs fault-independent search strategy introduced
in the redundancy identification algorithm FIRE~\cite{IyA96}. FIRE identifies undetectable faults which
require conflicting value assignments on a single line in the circuit
for their detection.  

Some other extensions of FIRE have been proposed.  In~\cite{GuH00},
conflicting value assignments for pairs of vertices rather than single
vertices are considered. In~\cite{Hs02}, a technique for maximizing
conflicting value assignments on a single vertex is presented. A
large number of direct and indirect logic implications are derived and
stored in an implication graph. These implications are used to
increase the implication power of FIRE.  In~\cite{ViSH05}, binary
resolution in addition to static logic implications are used for
maximizing conflicting value assignments on multiple vertices. All
approaches described above allow for identification of more
undetectable faults compared to FIRE, but they make the complexity 
prohibitive for large circuits.

The presented algorithm differs from FIRE in several aspects. The first 
improvement  is an increased
implication power.  A fundamental difference of the presented approach
from other extensions of FIRE~\cite{GuH00,Hs02,ViSH05} is that we do
not perform any extra search to find additional implications.  Rather,
we re-use the information which is anyway available in the algorithm's flow.

\begin{figure}[t!]
\begin{center}
\includegraphics[width=0.67\columnwidth]{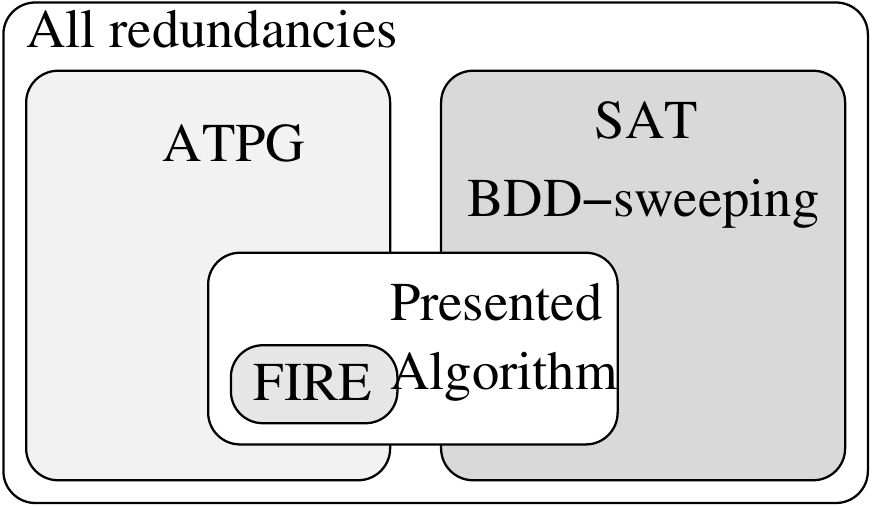}
\caption{Redundancy identification coverage of different methods.} \label{f0}
\end{center}
\end{figure}

The second improvement is the ability to identify some vertices which
implement equivalent or complemented functions.  Similarly to the
previous improvement, this is done with a minimum search, by re-using
the information from the algorithm's flow.  This improvement allows us to find
some redundancies which cannot be found by an ATPG-based approach 
(see Figure~\ref{f0}). To our best knowledge, the presented technique
is the first which can find structurally different equivalent vertices
in a circuit in less than exponential time.

The runtime improvements include a reduced number of unobservability
checks during unobservability propagation stage and a special
treatment of vertices with a single input.  Overall, the proposed
improvements allow us to find 37\% more redundancies than FIRE without
increasing its runtime.

Another difference from FIRE is that the presented algorithm removes redundancy,
while FIRE is only an identification algorithm.  The fundamental
problem of redundancy removal is to keep implication database updated
after a net or gate has been removed from the circuit. In the worst
case, the complete database has to be re-calculated.  We
use properties of indirect implications which allow us to update the
implications database instantly.

The paper is organized as follows. Section~\ref{nota} introduces the
notation and definitions used in the sequel.  Section~\ref{prelim} gives the background
on FIRE algorithm. Section~\ref{alg} presents the new algorithm. Section~\ref{impr} describes in details new contributions and differences between our approach and the one of FIRE.
Section~\ref{exp} summarizes experimental results.  Section~\ref{con}
concludes the paper and discusses open problems.

\section{Notation and Definitions} \label{nota}

{\em Undetectable fault} is a fault which cannot be detected by any
input pattern~\cite{Fr67}.  This can happen either because it is not
possible to apply to the faulty line the value opposite to the value
of the fault ({\em uncontrollability}), or because the effect of the
fault cannot be propagated to the output ({\em unobservability}). In a
combinational circuit, undetectable stuck-at faults are always caused
by redundancy~\cite{IyA96}.

Let $C=(V,E)$ be a Boolean circuit, where $V$ represents gates and
primary inputs and $E$ describes the nets connecting the gates.  We
use the letters $v, u$ and $w$ to denote the vertices of $C$.  The
letters $s$ and $q$ are designated for the {\em stem} of a multiple
fanout net, and $b_1, \ldots, b_r$ for its {\em branches} (as in
Figure~\ref{ex1}).

The set of predecessors of a vertex $v \in V$ is denoted by $IN(v) =
\{u \in V \ | \ (u,v) \in E\}$.  The set of successors of $v$ is
denoted by $OUT(v) = \{u \in V \ | \ (v,u) \in E\}$.

A value on the input of a gate is {\em controlling}, if its presence
determines the value of the gate's output, independently of the values
of other inputs.
A value on the output of a gate is {\em controlled} if it was set by
a controlling input value.  For AND (OR) the controlling and controlled
values are the same, namely 0 (1).  For NAND (NOR) they are 0 (1) and
1 (0), respectively. The XOR has no controlling and controlled values. 

A logic implication is {\em direct} if it relates inputs and output of
a single gate and it is evident from the type of this gate only. For
example, 0(1) at one of the inputs of an AND(OR) directly implies 0(1)
on gate's output; 1(0) at the output of an AND(OR) directly implies
1(0) at all inputs of the gate.

\section{FIRE Algorithm}
\label{prelim}

FIRE algorithm~\cite{IyA96} classifies a stuck-at fault on the line $l$ as
undetectable if this fault requires the presence of both, 0 and 1
values (i.e. a conflict) on some other line $r$ as a necessary
condition for its detection. All stems in the circuit are checked as
candidates to be such a line $r$.  For each stem $q$, two sets of
faults, $set_0$ and $set_1$, are computed.  The set $set_i$ is defined
as the set of faults that require $q$ to have value $i \in \{0,1\}$ as
a necessary condition for their detection.  The set of undetectable
faults is obtained by intersecting $set_0$ and $set_1$.

\begin{figure}[t!]
\begin{center}
\includegraphics[width=0.5\columnwidth]{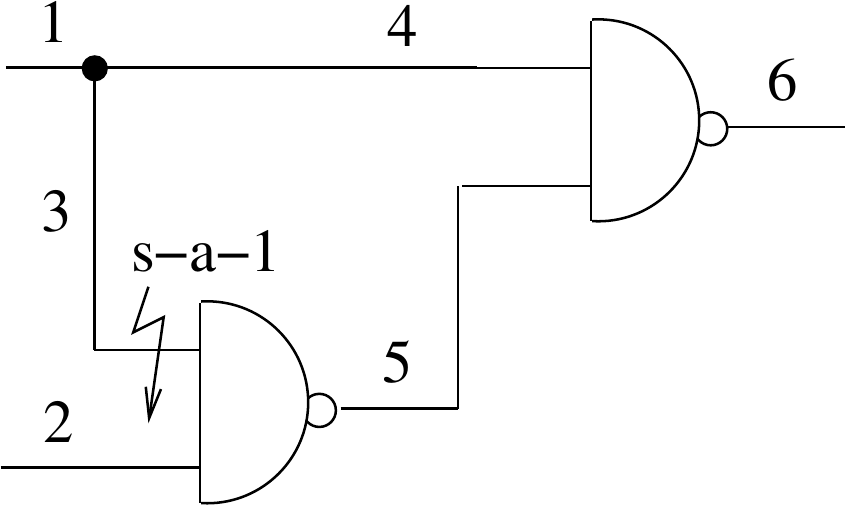}
\caption{An example of undetectable fault.}\label{ex}
\end{center}
\end{figure}

As an example, consider the circuit shown in
Figure~\ref{ex}. Stuck-at-1 fault on line 3 requires the value 1 on
line 1 for observability and the value 0 on line 1 for
controllability. Therefore, this fault in undetectable.

In order to compute $set_i$ for a stem $q$ the FIRE algorithm does the
following.  The value $\overline{i}$ is set on $q$ and constant
propagation using direct implications is applied recursively.  If some
line $l$ is assigned the value 1(0), it is uncontrollable for the
value 0(1), so stuck-at-1(0) fault at $l$ is added to $set_i$. This is
because the line $l$ cannot be assigned value 0(1) and thus cannot be
tested for stuck-at-1(0) fault when $q$ has the value $\overline{i}$.
Propagation of constants may result in some lines becoming
unobservable.  If one input of a gate is set to the controlling value,
then all other inputs of this gate become unobservable. If some line
$l$ is unobservable, then both stuck-at-0 and stuck-at-1 at $l$ are
added to $set_i$. The unobservability is propagated backward.  If all
fanout branches of a stem $s$ are marked unobservable, the following
Lemma is applied to decide whether $s$ is also unobservable or not.

\begin{lemma} \label{l1} \cite{IyA96} 
A stem $s$ with all its branches marked as unobservable may also
be marked as unobservable if, for each branch $b$ of $s$, there
exists at least one set of lines $\{l_b\}$ such that the following
conditions are satisfied:
\begin{enumerate}
\item the branch $b$ is unobservable because of uncontrollability 
indicators on every line in $\{l_b\}$, and
\item every line in  $\{l_b\}$ is unreachable from $s$.
\end{enumerate}
\end{lemma}

\begin{figure}[t!]
\begin{center}
\includegraphics[width=0.7\columnwidth]{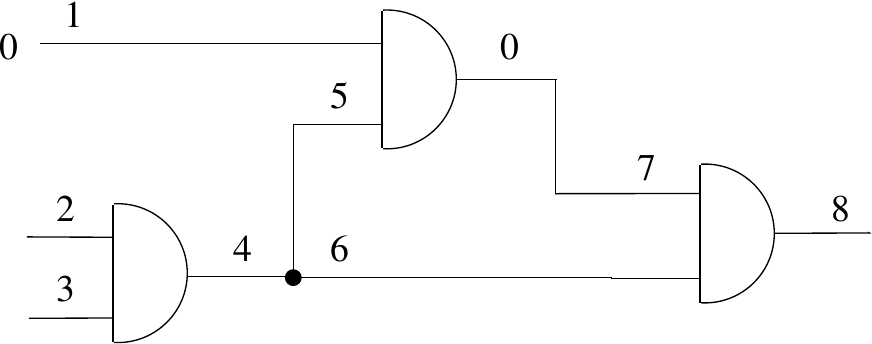}
\caption{Example showing that the conditions formulated in 
Lemma~\ref{l1} are sufficient but not necessary.} \label{f2}
\end{center}
\end{figure}

The conditions formulated in Lemma~\ref{l1} are sufficient, but not
necessary, conditions for unobservability. 
An example is shown in Figure~\ref{f2}.  Suppose that the
line 1 has the value 0. Then, both fanout branches of the net 4 are
unobservable. For the branch 5, the line 1 satisfies both conditions
of Lemma~\ref{l1}, since 5 is unobservable because of the 0 value on
the line 1, and 1 is unreachable from 4. The branch 6 is unobservable
because of the 0 value on the line 7, however, 7 is reachable from 4.
So, the second condition of Lemma~\ref{l1} does not hold and therefore
the net 4 is not recognized by FIRE as unobservable.

\section{Redundancy Removal}
\label{alg}

The overall flow of presented algorithm is similar to the one of FIRE.
The pseudo-code is shown in Figures~\ref{code}
and~\ref{code1}. In this section, we describe the implementation
details necessary for its understanding. In the Section~\ref{impr}, we
focus on the conceptual differences between the presented algorithm and FIRE.

In the main loop (steps 8 - 58), our algorithm iterates through all
vertices of the circuit, $v_{base} \in V$, sets them to the value $i$,
$i \in \{0,1\}$, and performs uncontrollability and unobservability
propagation in order to compute the information similar to $set_i$ for
the stem fed by $v_{base}$ in FIRE algorithm.  We use the term {\em
$i$th run for} $v_{base}$ to refer to the uncontrollability and
unobservability propagation with the vertex $v_{base}$ being set to
the value $i$, $i \in \{0,1\}$.

In our implementation, every vertex $v \in V$ has the following
fields:

\begin{enumerate}
\item  $unobservable\_outputs(v,i) \in \{0,1,\ldots,|OUT(v)|\}$ is 
the number of outgoing edges of $v$ which become unobservable after
$i$th run.
\item $master(v,i) \subset IN(v) \cup \{NULL, ALL\}$ is
a pointer to the predecessor of $v$ which has the controlling value
for $v$ in the $i$th run. Also, $master(v,i) = u$ means that all
inputs of $v$, except $u$, are unobservable.  If there is more than
one such predecessor, $master(v,i)$ is set to the dummy vertex $ALL$,
which means that all inputs of $v$ are unobservable. Initially, when
vertices are not set to values, $master(v,i)$ is set to $NULL$, which
means that no input of $v$ is unobservable.
\item $visited(v) \in \{0,1\}$ shows whether $v$
has been visited or not during unobservability propagation and check,
$visited(v) = 1$ if visited, 0 otherwise.
\item $indirect\_implications(v,j)$, $j \in \{0,1\}$, contains the 
list of pairs of type $(u,k)$, $u \in V$, $k \in \{0,1\}$, such that
there exists a logic implication $(v = j) \rightarrow (u = k)$.
\item $invalid\_implication(u) \in \{on,off\}$ shows whether 
all indirect implications which imply $u$ to some value are valid or
not.  If $invalid\_implication(u) = on$, then all implications of type
$(v = j) \rightarrow (u = k)$ stored as $(u,k) \in
indirect\_implications(v,j)$ for some $v \in V$, $j,k \in \{0,1\}$,
are not valid.
\end{enumerate}

For each $v_{base}$, the presented algorithm first performs constant propagation and
indirect implications learning (steps 13-22, explained in subsection
5.3), then eliminates duplicated and constant vertices (steps 23-40,
explained in subsection 5.4), does unobservability propagation
(steps 41-46) and finally removes the
identified redundancies (steps 47-55).
 
The procedure $\procname{PropagateUncontrolability}$ performs constant
propagation, i.e. it recursively applies direct and learned indirect
implications following from the assignment $v_{base} = i$, $i \in
\{0,1\}$. The obtained values are stored not only with vertices, but
also in the queue $Q(i)$ defined by
\[
Q(i) := \{ (u,j) \ | \ u = j \ \mbox{after $i$th run for} \ v_{base} \}.
\]
If the justification of $v_{base}$ to $i$ causes a contradiction 
(i.e some vertex needs to be assigned to different values), then
$\procname{PropagateUncontrollability}$ returns $\O$.
$\procname{PropagateUncontrolability}$ also fills the field
$master(v,i)$ for all gates $v \in V$ in correspondence with its
definition.

The procedure $\procname{OverapproximateUnobservability}$
$\procname{Init}(v,i)$ initiates the process of unobservability
propagation.  It is invoked at all vertices which are set to
controlled values.  The procedure
$\procname{CheckUnobservability}(v,u,i)$ checks whether a given edge
$(v,u)$ is really unobservable or not. Specific features of unobservability
propagation and checking are discussed in subsection 5.1.

\section{Improvements over FIRE}
\label{impr}

In this section, we describe the improvements of the presented algorithm over FIRE.
The first two are runtime improvements, the second two are quality
improvements.

\subsection{Runtime Improvements}

\subsubsection{Reducing the number of unobservability checks during
unobservability propagation} \label{r1}

The first improvement in runtime is achieved by reducing number of
unobservability checks during unobservability propagation stage.

The initial source of unobservability in the circuit are vertices
which have at least one input set to a controlling value.  If an input
of a vertex $v$ has a controlling value, other inputs of $v$ become
unobservable. The rest of unobservable lines is derived by
unobservability propagation.

\begin{figure}[t!]
\begin{center}
\includegraphics[width=0.55\columnwidth]{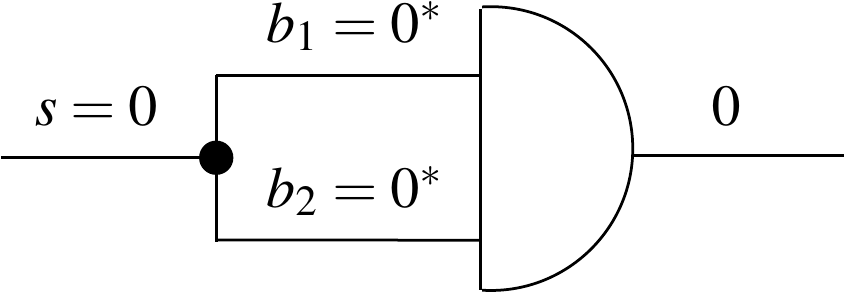}
\caption{An example showing that 
unobservability of all fanout branches does not necessarily imply
unobservability of the stem; the sign $''*''$ indicates that the value
is unobservable.} \label{ex1}
\end{center}
\end{figure}

Usually, if all fanout branches of a stem $s$ are unobservable, $s$ is
unobservable as well. However, there are some rare exceptions.  For
example, consider the circuit shown in Figure~\ref{ex1}. Suppose the
stem $s$ is set to 0 value.  Then, both branches $b_1$ and $b_2$ are
unobservable. In spite of this, $s$ is observable.

FIRE resolves the problem with such cases by checking unobservability
of a stem $s$ any time all its branches are identified as
unobservable (by applying Lemma~\ref{l1}).  Since
the unobservability check may need to be done multiple times per each
run, and number of runs is the number of vertices times two, the
number of unobservability checks made by FIRE equals to the number of
vertices multiplied by some factor $m$, which represents the average
number of unobservability checks per each pair of runs:
\begin{equation}
N_{unobservability\_checks}^{FIRE} = |V| \times m.
\end{equation}

Even though the factor $m$ can be up to the number of stems in the
circuit, our experience is that $m$ is usually small, although larger
than 1 in most of the cases.

In the presented algorithm, if we encounter a stem $s$ with all branches being
unobservable, we assume unobservability of $s$ without any check,
i.e. we overapproximate actual unobservability.  The eventual
correction is postponed for later.  If no redundancy is identified
with the overapproximated unobservability, no correction is performed
because this unobservability is not going to be used for redundancy
removal anyway. This differs our approach from FIRE, where the correction is
done always.

The number of unobservability checks needed for the presented algorithm equals
to the number of redundancies in the circuit plus the number of
incorrectly detected redundancies (caused by the overapproximated
unobservability):
\begin{equation} \label{eqr}
N_{unobservability\_checks} = N_{redundancies} + N_{incorrect\_redundancies}.
\end{equation}

In practice, we very rarely have cases where stems are not
unobservable if all their branches are unobservable.  Thus, we have
very few incorrectly identified unobservabilities, and, as a
consequence, very few incorrectly detected redundancies. Furthermore,
the number of redundancies in a circuit is usually quite
small. 

We further reduce the number of unobservability checks by 
using the following property. It shows that a line identified as
unobservable by the overapproximation procedure does not need to be checked
if it is not set to a constant value.

\begin{theorem} \label{th1}
If during the $i$th run, $i \in
\{0,1\}$, 
a line $l$ is classified as unobservable by the overapproximation
procedure and it is not set to a constant value,
then $l$ is unobservable.
\end{theorem}

\begin{figure}[t!]
\begin{center} 
{\scriptsize
\begin{tabbing} 
m \== m \== m \== m \== m \== m \== m \== m \== m \== \kill
\ALGORITHM $\procname{Redundancy Removal}(V,E)$ \\ 
\> Boolean $i, j$; \\
\> Vertex  $v, u, w, v_{base}$; \\
\> Queue   $Q(0), Q(1)$; \ /* lists of pairs $($Vertex $v$, Boolean $i)$ */ \\ 
\> integer $index$, $p$; \\ \\

1. \> $index := 0;$ \\
2. \> \FOR \ every vertex $v \in V$ \ in forward topological order \DO \\ 
3. \> \> $index(v) := index;$ ~$index++$; \\
4. \> \> $indirect\_implications(v,0) := \O$; \\
5. \> \> $indirect\_implications(v,1) := \O$; \\
6. \> \> $invalid\_implications(v) := off$; \\
7. \> \END \FOR 2 \\

8. \> \FOR \ $p$ from 0 to $|V|$ \DO \\
9. \> \> $v_{base}$ is a vertex with $index(v_{base})= p$; \\ 
10. \> \> \IF \ $|IN(v_{base})| = 1$ \CONTINUE \ /* Runtime improvement 2 (Section~\ref{r2}) */ \\

11. \> \> \BL: \\
12. \> \> For every $v \in V$, set $unobservable\_outputs(v,i)=0$,$master(v,i)=NULL$; \\ 
/* Uncontrollability propagation */ \\
13. \> \> \FOR \ every $i \in  \{0,1\}$ \ \DO \\ 
14. \> \> \> $Q(i) = \procname{PropagateUncontrolability}(v_{base},i)$; \\
15. \> \> \> \IF  $Q(i)$ is empty \ \THEN /* could not justify $v_{base}$ to $i$ */ \\
\> \> \> \> /* stuck-at-$\overline{i}$ at the output of  $v_{base}$ is undetectable */ \\
16. \> \> \> \> $\procname{UpdateImplications}(v_{base}, v_{base})$; \\
17. \> \> \> \> replace $v_{base}$ by the constant $\overline{i}$; \\
18.  \> \> \> \> \GO \TO \EL 57; \\

19. \> \> \> \FOR \ every $(v, \overline j) \in Q(i)$ \ \DO \\ 
20. \> \> \> \> add $(v_{base}, \overline i)$ to $indirect\_implications(v,j)$; \\
21. \> \> \> \END \FOR 19 \\

22. \> \> \END \FOR 13 \\

/* Removal of constant vertices. Quality improvement 2 (Section~\ref{q2}) */ \\
23. \> \> \FOR \ every pair $(v,j) \in Q(0) \cap Q(1)$ \ \DO \\
\> \> \> /* stuck-at-$j$ at at the output of  $v$ is undetectable */ \\
24. \> \> \> $\procname{UpdateImplications}(v, v_{base})$; \\
25. \> \> \> replace $v$ by the constant $j$; \\
26. \> \> \END \FOR 23 \\ 

/* Removal of duplicated vertices. Quality improvement 2 (Section~\ref{q2}) */ \\
27. \> \> Create $\overline Q(1): = \{(v,i) | (v,\overline i) \in Q(1) \}$; \\ 
\> \> /* $\overline Q(1)$ contains all pairs of $Q(1)$ with $i$ being complemented */ \\
28. \> \> \FOR \ every pair $(u,j) \in \overline Q(1) \cap Q(0)$ \ \DO \\
29. \> \> \> \IF \ $i = 0$ \ \THEN \\
30. \> \> \> \> Add $u$ to the equivalence class of $v_{base}$, $E(v_{base})$; \\
31. \> \> \> \ELSE \\
32. \> \> \> \> Add $u$ to the equivalence class of $\overline v_{base}$, $E(\overline v_{base})$ \\
33. \> \> \END \FOR 28 \\ 
34. \> \> Replace all $v \in E(v_{base})$ by $u \in E(v_{base})$ closest to primary inputs; \\
35. \> \> Replace all $v \in E(\overline v_{base})$ by $w \in E(\overline v_{base})$ 
closest to primary inputs; \\

36. \> \> \IF $|IN(u)| = 1$ and $|IN(w)| = 1$ \THEN \\
37. \> \> \> \IF $u$ is closer to primary inputs than $w$ \THEN \\
38. \> \> \> \> Substitute $w$ by an inverter fed by $u$; \\
39. \> \> \> \ELSE \\
40. \> \> \> \> Substitute $u$ by an inverter fed by $w$; \\

/* Unobservability propagation */ \\
41. \> \> \FOR \ every $i \in  \{0,1\}$ \ \DO \\ 
42. \> \> For every  $v \in V$, set $visited(v)=0$; \\
43. \> \> \> \FOR \ every pair $(v,j) \in Q(i)$ \ \DO \\
44. \> \> \> \> \IF \ $j$ is the controlled value of $v$ \ \THEN \\
45. \> \> \> \> \> $\procname{OverapproximateUnobservabilityInit}(v,i)$; \\
46. \> \> \> \END \FOR 43 \\

/* Search for a redundant line */ \\
47. \> \> \> \FOR \ every $(v,j) \in Q(\overline{i})$ \ such that \ $unobservable\_outputs(v,i) > 0$ \ \DO \\
48.  \> \> \> \> \FOR \ every $u \in OUT(v)$ \ \DO \\
49. \> \> \> \> \> \IF \ $master(u,i) \not= NULL$ \AND $master(u,i) \not= v$ \ \THEN \\

50. \> \> \> \> \> \> \IF \ the value of $v$ is not set to a constant during $i$th run \OR \\
 \> \> \> \> \> \> ~~~ $\procname{CheckUnobservability}(v,u,i) = \true$ \ \THEN \\
\> \> \> \> \> \> \> /* stuck-at-$j$ at $(v,u)$ is undetectable */ \\
51. \> \> \> \> \> \> \>  $\procname{UpdateImplications}(u, v_{base})$; \\
52. \> \> \> \> \> \> \> replace $(v,u)$ by the constant $j$; \\
53. \> \> \> \> \> \> \> \GO \TO \BL 11; \\

54. \> \> \> \> \END \FOR 48 \\
55. \> \> \> \END \FOR 47 \\ 
56. \> \> \END \FOR 41 \\ 
57. \> \> \EL: \\
58. \> ~~~~ \END \FOR 8 \\ 
59. \END 
\end{tabbing}
}
\caption{Pseudo-code of the presented algorithm.}\label{code} 
\end{center}
\end{figure}

\begin{figure}[t!]
\begin{center} 
{\scriptsize
\begin{tabbing} 
m \== m \== m \== m \== m \== m \== m \== m \== m \== \kill
\ALGORITHM $\procname{OverapproximateUnobservabilityInit}(v,i)$ \\ 
\> \IF \ $visited(v) = 0$; \\ 
\> \> $visited(v) := 1$; \\
\> \> \FOR \ every vertex $u \in IN(v) - \{master(v,i)\}$  \ \DO \\ 
\> \> \> $\procname{CheckOutputs}(u,i)$ \\
\> \> \END \FOR  \\
\END \\ \\ \\

\ALGORITHM $\procname{CheckOutputs}(v,i)$ \\
\> $unobservable\_outputs(v,i)$++; \\
\> \IF $unobservable\_outputs(v,i) = |OUT(v)|$ \ \THEN \\
\> /* Runtime improvement 1 (Section~\ref{r1}). Here FIRE does */\\
\> /* unobservability check, while the presented algorithm propagates */\\
\> /* unobservability without any check */ \\
\> \> $master(v,i) := ALL$; \\
\> \> $\procname{OverapproximateUnobservability}(v,i)$; \\
\END \\ \\ \\

\ALGORITHM $\procname{OverapproximateUnobservability}(v,i)$ \\ 
\> \IF \ $visited(v) = 1$; \\ 
\> \> $\procname{CheckOutputs}(master(v,i),i)$ \\
\> \ELSE \ \\
\> \> $visited(v) := 1$; \\
\> \> \FOR \ every vertex $u \in IN(v)$  \ \DO \\ 
\> \> \> $\procname{CheckOutputs}(u,i)$ \\
\> \> \END \FOR  \\
\END \\ \\ \\

\ALGORITHM $\procname{CheckUnobservability}(v,u,i)$ \\ 
\> For every vertex $w \in V$, set $visited(w)=0$; \\
\> \IF $\not\exists w \in IN(u)-\{v\}$ such that
value of $w$ in the $i$th run is controlling for $u$ \\ 
\> \> \IF \ $unobservable\_outputs(u,i) \not= |OUT(u)|$ \ \THEN \\
\>  \> \> \RETURN \ $\false$; \\
\> \> \ELSE \\
\> \> \> \IF \ $\procname{Unobservability}(u,i) = \false$ \ \THEN \\
\> \> \> \> \RETURN \ $\false$; \\
\> \RETURN \ $\true$; \\
\END \\ \\ \\

\ALGORITHM $\procname{Unobservability}(v,i)$ \\ 
\> \IF \ $visited(v) = 1$ \ \THEN \\
\> \> \RETURN \ $\true$; \\

\> $visited(v) := 1$; \\
\> \FOR \ every $u \in OUT(v)$ \ \DO \\
\> \> \IF $\not\exists w \in IN(u)$ such that $visited(w) \not= 1$ \AND \\
\> \> ~~~ value of $w$ in the $i$th run is controlling value for $u$ \ \THEN \\
\> \> \> \IF \ $unobservable\_outputs(u,i) \not= |OUT(u)|$ \ \THEN \\
\> \> \> \> \RETURN \ $\false$; \\
\> \> \> \ELSE \\
\> \> \> \> \IF \ $\procname{Unobservability}(u,i) = \false$ \ \THEN \\
\> \> \> \> \> \RETURN \ $\false$; \\
\> \END \FOR \\
\> \RETURN \ $\true$; \\
\END \\ \\ 

\ALGORITHM $\procname{UpdateImplications}(v, v_{base})$ \\ 
\> /* Quality improvement 1 (Section~\ref{q1}) */ \\
\> $\procname{UpdateR1}(v)$; /* updating region $R_1$ */ \\
\> \FOR \ every $u$ such that $index(v) < index(u) < index(v_{base})$ \\
\> \> $invalid\_implication(u) := on$; /* updating region $R_2$ */  \\
\> \END \FOR \\
\END \\ \\

\ALGORITHM $\procname{UpdateR1}(v)$ \\ 
\> \IF \ $indirect\_implication(v,0) \not= \O$ \ OR $indirect\_implication(v,1) \not= \O$ \ \THEN \\
\> \> $indirect\_implication(v,0) := \O$; \\
\> \> $indirect\_implication(v,1) := \O$; \\
\> \> \FOR \ every $u \in OUT(v)$ \ \DO \\
\> \> \> $\procname{UpdateR1}(u)$; \\
\> \> \END \FOR \\
\END 
\end{tabbing}
}
\caption{Pseudo-codes of the procedures.}\label{code1} 
\end{center}
\end{figure}




The Theorem~\ref{th1} is used at the step 50 of the pseudocode, where
the algorithm checks whether the value of $v$ is not set to a constant
during the $i$th run. Only if this statement is not satisfied, the
unobservability check $\procname{CheckUnobservability}(v,u,i)$
is invoked.  Note that, in order to reach the step 50, the
conditions of {\bf for}-loop at the step 47 need to be satisfied.  As a
consequence, the unobservability check is done only if the line
$(v,u)$ is set to a constant value in both runs.  This is possible
only if the gate $v$ is either a constant function, or it is
equivalent to $v_{base}$ or $\overline{v}_{base}$.  Since constant and
duplicated functions are removed earlier (at the steps 23-40),
unobservability checks have to be performed only if some output of
$v_{base}$ or $\overline{v}_{base}$ was identified as redundant.
Thus, terms $N_{redundancies}$ and $N_{incorrect\_redundancies}$ in
the equation~\ref{eqr} refer not to all identified redundancies, but
rather to the redundancies found at the output of $v_{base}$ or
$\overline{v}_{base}$.  Therefore, the overall number of
unobservability checks of the presented algorithm is significantly smaller than the
one of FIRE, while the complexity and average runtime of each check is
the same for both algorithms.

\subsubsection{Special treatment of gates with a single input} \label{r2}

The second improvement in runtime is a result of skipping the outer
{\bf for}-loop for the vertices with a single input.  All implications
which could be obtained during these runs would be equivalent to the
ones found during the runs of the algorithm for the only predecessor of
the vertex.  Therefore, no new redundancies would be identified.

\subsection{Quality Improvements}

\subsubsection{Increased implication power} \label{q1}

The first improvement in quality is achieved by re-using the
information from the previous runs of the algorithm to increase its
implication power in the future runs. This is done as follows.

\begin{figure}[t!]
\begin{center}
\includegraphics[width=0.7\columnwidth]{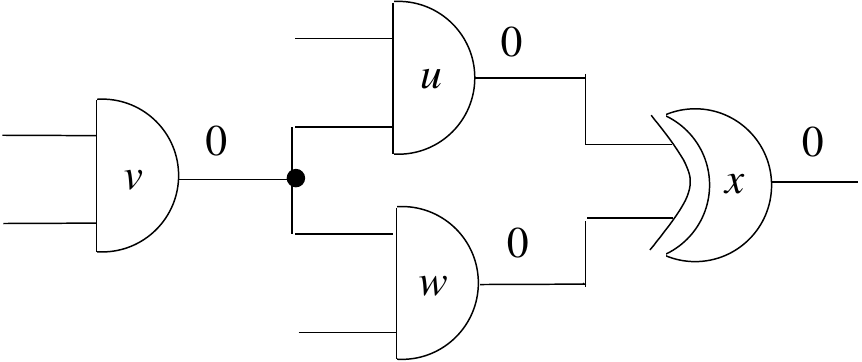}
\caption{An example showing how an indirect implication can be obtained.} \label{ex2}
\end{center}
\end{figure}

Suppose that, during $i$th run for the vertex $v_{base}$,
the vertices $u_1,
\ldots, u_p$ are set to some values $j_1, \ldots, j_p$.  We can conclude that
the assignment of $v_{base} = i$ implies the values $j_1, \ldots, j_p$
on $u_1,
\ldots, u_p$. From this, by using the contrapositive low, we can
derive a set of logic implications $(u_1 = \overline j_1) \rightarrow
(v_{base} = \overline i), \ldots, (u_p = \overline j_p) \rightarrow (v_{base} =
\overline i)$.  Some of them might be indirect implications.  For
every $r \in \{1,\ldots,p\}$, we store the implication $(u_r =
\overline j_r) \rightarrow (v_{base} = \overline i)$ in the field
$indirect\_implications$ of $u_i$ by adding to the field the pair
$(v_{base},\overline i)$.  When we justify the vertex $u_r$ to the
value $j_r$ in later runs, these stored indirect implications
will be used to set the vertex $v$ to the value $k$, for every $(v,k)
\in indirect\_implication(u_r,j)$, provided $invalid\_implication(v) =
off$.

Note, that indirect implications are not immediately evident from the
circuit. They cannot always be derived by FIRE approach. As an
example, consider the circuit shown in Figure~\ref{ex2}. During 0th run
for the vertex $v$, the vertices $u, w$ and $x$ are set to 0 by constant
propagation. By using the contrapositive low, we derive logic
implications $(x = 1) \rightarrow (v = 1), (u = 1) \rightarrow (v = 1)$
and $(w = 1) \rightarrow (v = 1)$.  While the last two are direct, the
first one is not. FIRE algorithm would not set the vertex $v$ to 1 when
$x$ is justified to 1.

A fundamental difference of the presented algorithm from other approaches which
improve FIRE by increasing its implication
power~\cite{GuH00,Hs02,ViSH05} is that we do not perform any extra
search to find indirect implications.  Rather, we re-use information
available from the algorithm's flow.

Another important contribution is the following observation. Every
time a redundant line is removed, some of the indirect implications
become invalid.  On one hand, checking whether each indirect
implication needs to be removed or not would be expensive.  On the
other hand, erasing all indirect implications would be a waste of many
implications which are still valid.  We found a property of indirect
implications in our algorithm which localizes the region where invalid
implications may appear. Thus, only this region needs to be cleaned.

Let $l$ be a redundant line which is found in one of the runs for the
vertex $v_{base}$. We define two regions, $R_1, R_2  \subseteq V$, as follows.
$R_1$ consists of all vertices which can be reached from
$l$ without passing through any vertex $w$ with no indirect
implication, i.e. such that $indirect\_implication(w,0) = \O$ and
$indirect\_implication(w,1) = \O$.
$R_2$ is empty if $index(v_{base}) < index(u)$, where $u$ is the
vertex fed by $l$.  Otherwise, $R_2$ consists of all vertices whose
indexes are in the interval between $index(u)$ and $index(v_{base})$.

\begin{theorem}
An indirect implication $(v=i) \rightarrow (u=j)$ can be invalid only if
either $v \in R_1$, or $u \in R_2$, for some $i,j \in \{0,1\}$.  
\end{theorem}

It follows from the above theorem that, in order to make the implications
consistent, it is sufficient to visit every vertex $v \in R_1$ and
remove all implications stored with $v$, i.e. set
$indirect\_implication(v,j) = \O$ for all $j \in \{0,1\}$.  In addition, 
set $invalid\_implication(u) = on$ for all $u \in R_2$.
The updating is done by the procedure $\procname{UpdateImplications}$.

Our experience is that indirect implications usually do not spread
many levels forward from the level of currently processed vertex
$v_{base}$. Therefore, $R_1$ is quite small. Normally, it has a
constant size regardless of the circuit size.  The region $R_2$ is
almost always empty, because $index(u)$ is less than $index(v_{base})$
for 1 out of 1000 found redundancies on average. As a result, the
updating process is very quick.

Note, that not all implications which we remove are invalid.  However,
checking their validity would take a considerable amount of time, since
 we do not keep track of how they were found. So, by removing them all, we
save time. Furthermore, since $R_1$ and $R_2$ are small, 
most of valid indirect implications are left after updating.

\subsubsection{Identification of duplicated functions} \label{q2}

The presented algorithm is able to identity some vertices which implement equivalent
or complemented functions, as well as constant vertices.  Therefore,
it can remove some redundancies which cannot be found by FIRE or ATPG.

Equivalent functions are identified as follows.
If the assignment of the vertex $v_{base}$ to the value $i$ causes some gate
$u$ to be set to the value $i$, as well as the assignment of $v_{base}$ to
$\overline i$ causes $u$ to be set to $\overline i$, then
we conclude that $v$ and $u$ implement the same function and
add $u$ to the equivalence class of $v_{base}$, $E(v_{base})$.

Similarly, to identify complemented functions, 
$(v_{base}=i) \rightarrow (u=\overline i)$ and
$(v_{base}=\overline i) \rightarrow (u=i)$ imply that $v_{base}$ and
$u$ are complements of each other.  The vertex $u$ is added to the equivalence
class of $\overline v_{base}$, $E(\overline v_{base})$.

Finally, all vertices in $E(v_{base})$ and $E(\overline v_{base})$ are
replaced by a member of the class which is closest to the primary
inputs in topological order (i.e. has the shortest longest path to the
primary inputs). Let $u$ be the member selected from $E(v_{base})$ and
$w$ be the member selected from of $E(\overline v_{base})$. If both
$u$ and $w$ have more than one input, then the one which is further
away from the primary inputs is substituted by an inverter fed by the
other one.

To identify constant functions, we use the following simple property.
If $(v_{base}=0) \rightarrow (u = i)$ and $(v_{base}=1) \rightarrow (u
= i)$, then the vertex $u$ is the constant $i$.

Note, that redundant vertices are identified with a minimum search, by
re-using the information available from the algorithm's flow. 

\begin{table*}[t!]\centering\footnotesize
\begin{tabular}{|c||c|c|c||c|c|c||c|c|c||c|c|} \hline
& \multicolumn{3}{c||}{} & \multicolumn{3}{c||}{Presented without} & \multicolumn{3}{c||}{} & \multicolumn{2}{c|}{} \\ 
& \multicolumn{3}{c||}{FIRE (results from~\cite{IyA96})} & \multicolumn{3}{c||}{4 improvements from Section~\ref{impr}} & \multicolumn{3}{c||}{Presented} 
& \multicolumn{2}{c|}{ATPG~\cite{BrD94}} \\ \cline{2-12}

name  & \# red & \% red & $t$, sec & \# red & \% red & $t$, sec & \# red & \% red & $t$, sec  & \# red 
& $t$, sec \\ \hline

C17   & -  & -    & -    & 0   & -    & 0.01 & 0   & -    & 0.00 & 0   & 0.00 \\ \hline
C432  & -  & -    & -    & 45  & 60.8 & 0.01 & 63  & 85.1 & 0.00 & 74  & 4.43 \\ \hline
C499  & -  & -    & -    & 0   & 0    & 0.01 & 0   & 0    & 0.01 & 8   & 0.27 \\ \hline
C880  & -  & -    & -    & 0   & 0    & 0.01 & 0   & 0    & 0.01 & 8   & 0.08 \\ \hline
C1355 & -  & -    & -    & 0   & -    & 0.02 & 0   & -    & 0.01 & 0   & 0.82 \\ \hline
C1908 & 6  & 15.4 & 1.8  & 24  & 61.5 & 0.02 & 26  & 66.7 & 0.02 & 39  & 0.51 \\ \hline
C2670 & 29 & 16.3 & 1.5  & 56  & 31.5 & 0.03 & 64  & 36.0 & 0.03 & 178 & 0.53 \\ \hline
C3540 & 93 & 38   & 11.9 & 144 & 58.8 & 0.08 & 167 & 67.9 & 0.09 & 246 & 2.17 \\ \hline
C5315 & 20 & 14   & 2.8  & 36  & 25.2 & 0.05 & 62  & 43.4 & 0.05 & 143 & 1.45 \\ \hline
C6288 & 33 & 82.5 & 1.3  & 69  & 98.6 & 0.06 & 113 & 161.4 & 0.06 & 70 & 2.00 \\ \hline
C7552 & 30 & 7.39 & 4.7  & 99  & 24.3 & 0.10 & 124 & 30.5 & 0.11 & 406 & 5.25 \\ \hline \hline
{\bf average } & 35.2 & 28.9 & 4.00 & 43 & 40.1 & 0.036 & 56.3 & 54.6 & 0.037 & 105.5 & 1.59 \\ \hline
\end{tabular}
\caption{Benchmark results for ISCAS'85 circuits; average is computed 
for non-``$-$'' entries.} \label{t1}
\end{table*}

\section{Experimental Results}
\label{exp}

This section compares the performance of the presented algorithm to
the ATPG-based approach from~\cite{BrD94}.  We also show the results
of FIRE~\cite{IyA96} as a reference. Note that FIRE only identifies
redundancy, but does not remove it.

Table~\ref{t1} summarizes the results for ISCAS'85 benchmark set.
Columns 2-4 show the results for FIRE, taken from~\cite{IyA96}. Column
2, \# red, is the total number of identified redundant lines. The sign
`-'' means that no result is reported in~\cite{IyA96}.  Column 3, \%
red, is the percentage of identified redundant lines compared to
ATPG~\cite{BrD94}, and Column 4 is CPU time, in seconds. According
to~\cite{IyA96}, FIRE was run on a SUN SPARC2. No parameters of the
SUN SPARC2 machine are provided in~\cite{IyA96}, so a comparison of
CPU times of the two algorithms is hard to make.  Therefore, in order
to evaluate the effect of four improvements described in
Section~\ref{impr}, we re-run the presented algorithm with these improvements
switched off.  The presented algorithm without these improvements can be considered
as our re-implementation of FIRE.  The results are shown in Columns
5-7. As we can see, runtime improvements fully compensate the time
consumed by quality improvements. If runtime improvements are switched
off, while quality improvements are on, the presented becomes about 20\%
slower.

Columns 8-10 and 11-12 show the corresponding numbers for the presented algorithm and
ATPG~\cite{BrD94}. The experiments were run on a PC with Pentium III
750 MHz processor and 256 Mb memory.  On average, for ISCAS'85
benchmarks, the presented algorithm removes 54.6\% of ATPG redundancies using only
2.28\% of its runtime.

On a larger set of 183 circuits from IWLS'02 benchmarks set with the
average size of 780 gates and the maximum size of 25000 gates
The presented algorithm without four improvements removes 14\% of ATPG's
redundancies, while with improvements it removes 19.3\% of ATPG's redundancies,
both using 2.5\% of ATPG's time. These results do not include two
benchmarks from the set, $spla$ and $pdc$, for which ATPG did not
finish in 2 hours, while the presented algorithm finished in 15 sec. As we can see,
the proposed improvements allow us to find 37\% more redundancies
without increasing the runtime.

\section{Conclusion}
\label{con}

This paper presents a heuristic algorithm which efficiently
identifies and removes redundancy in combinational circuits. Unlike
other extensions of FIRE, the presented algorithm provides better quality of results
without trading it for runtime. The speed of our heuristic makes it
suitable for running multiple times during synthesis. 

A possible extension of the presented approach is to employ an alternative strategy
for removing redundant lines. In our current implementation, we used
first-found first-removed approach in order to keep the runtime to
minimum.

\bibliographystyle{IEEEtran}
\bibliography{bib}

\begin{thebibliography}{10}
\providecommand{\url}[1]{#1}
\csname url@rmstyle\endcsname
\providecommand{\newblock}{\relax}
\providecommand{\bibinfo}[2]{#2}
\providecommand\BIBentrySTDinterwordspacing{\spaceskip=0pt\relax}
\providecommand\BIBentryALTinterwordstretchfactor{4}
\providecommand\BIBentryALTinterwordspacing{\spaceskip=\fontdimen2\font plus
\BIBentryALTinterwordstretchfactor\fontdimen3\font minus
  \fontdimen4\font\relax}
\providecommand\BIBforeignlanguage[2]{{%
\expandafter\ifx\csname l@#1\endcsname\relax
\typeout{** WARNING: IEEEtran.bst: No hyphenation pattern has been}%
\typeout{** loaded for the language `#1'. Using the pattern for}%
\typeout{** the default language instead.}%
\else
\language=\csname l@#1\endcsname
\fi
#2}}

\bibitem{BrM82}
R.~K. Brayton and C.~McMullen, ``The decomposition and factorization of
  {B}oolean expression,'' in \emph{Proceedings of the IEEE International
  Symposium of Circuits and Systems}.\hskip 1em plus 0.5em minus 0.4em\relax
  IEEE, 1982, pp. 49--54.

\bibitem{darringer}
J.~Darringer, W.~Joyner, L.~Berman, and L.~Trevillyan, ``Logic synthesis
  through local transformations,'' \emph{IBM Journal on Research and
  Development}, vol.~25, no.~4, pp. 272--280, July 1981.

\bibitem{SeD01}
E.~Sentovich and D.~Brand, \emph{Flexibillity in logic}.\hskip 1em plus 0.5em
  minus 0.4em\relax Norwell, MA, USA: Kluwer Academic Publishers, 2002.

\bibitem{LiLC95}
H.-C. Liang, C.~L. Lee, and J.~E. Chen, ``Identifying untestable faults in
  sequential circuits,'' \emph{IEEE Des. Test}, vol.~12, no.~3, pp. 14--23,
  1995.

\bibitem{AbI92}
M.~Abramovici and M.~A. Iyer, ``One-pass redundancy identification and
  removal,'' in \emph{Proceedings of the IEEE International Test Conference on
  Discover the New World of Test and Design}.\hskip 1em plus 0.5em minus
  0.4em\relax Washington, DC, USA: IEEE Computer Society, 1992, pp. 807--815.

\bibitem{Fr67}
A.~D. Friedman, ``Fault detection in redundant circuits,'' \emph{IEEE
  Transactions on Electronic Computers}, vol. EC-16, pp. 99--100, February
  1967.

\bibitem{ScA89}
M.~H. Schulz and E.~Auth, ``Improved deterministic test pattern generationwith
  application to redundancy identification,'' \emph{IEEE Transactions on
  Computer-Aided Design}, vol.~8, pp. 811--815, July 1989.

\bibitem{BeE02}
M.~Berkelaar and K.~M. van Eijk, ``Efficient and effective redundancy removal
  for million-gate circuits,'' in \emph{Proceedings of the Design, Automation
  and Test in Europe Conference and Exhibition}.\hskip 1em plus 0.5em minus
  0.4em\relax IEEE Computer Society Press, 2002, p. 1088.

\bibitem{BrD94}
D.~Brand and R.~Damiano, ``In the driver's seat of {B}oole{D}ozer,'' in
  \emph{Proceedings of International Conference on Computer Design}.\hskip 1em
  plus 0.5em minus 0.4em\relax IEEE, October 1994, pp. 518--521.

\bibitem{HaM89}
M.~Harihara and P.~R. Menon, ``Identification of undetectable faults in
  combinational circuits,'' in \emph{Proceedings of International Conference on
  Computer Design}.\hskip 1em plus 0.5em minus 0.4em\relax IEEE, October 1989,
  pp. 290--293.

\bibitem{MeA92}
P.~R. Menon and H.~Ahuja, ``Redundancy removal and simplification of
  combinational circuits,'' in \emph{Proceedings of VLSI Test Symposium}.\hskip
  1em plus 0.5em minus 0.4em\relax IEEE, April 1992, pp. 268--273.

\bibitem{IyA96}
M.~A. Iyer and M.~Abramovici, ``{FIRE}: A fault-independent combinational
  redundancy identification algorithm,'' \emph{IEEE Transactions on VLSI
  Systems}, vol. EC-16, pp. 295--301, June 1996.

\bibitem{GuH00}
K.Gulrajani and M.~S. Hsiao, ``Multi-node static logic implications for
  redundancy identification,'' in \emph{Proceedings of Design, Automation and
  Test in Europe Conference}.\hskip 1em plus 0.5em minus 0.4em\relax IEEE,
  March 2000, pp. 729--735.

\bibitem{Hs02}
M.~S. Hsiao, ``Maximizing impossibilities for untestable fault
  identification,'' in \emph{Proceedings of Design, Automation and Test in
  Europe Conference}.\hskip 1em plus 0.5em minus 0.4em\relax IEEE, March 2002,
  p. 949.

\bibitem{ViSH05}
V.~C. Vimjam, M.~Syal, and M.~S. Hsiao, ``Testing: Untestable fault
  identification through enhanced necessary value assignments,'' in
  \emph{Proceedings of the 15th ACM Great Lakes symposium on VLSI}.\hskip 1em
  plus 0.5em minus 0.4em\relax IEEE, 2005.

\bibitem{KiSSS97}
J.~P. Kim, M.~Silva, H.~Savoj, and K.~A. Sakallah, ``{RID-GRASP}: Redundancy
  identification and removal using {GRASP},'' in \emph{Proceedings of
  International Workshop on Logic Synthesis}.\hskip 1em plus 0.5em minus
  0.4em\relax IEEE, May 1987.

\bibitem{kuehlmann}
A.~Kuehlmann, M.~Ganai, and V.~Paruthi, ``Robust {B}oolean reasoning for
  equivalence checking and functional property verification,''
  \emph{Transactions on Computer-Aided Design of Integrated Circuits and
  Systems}, vol.~21, no.~12, pp. 1377--1394, December 2002.

\bibitem{KuKr97}
A.~Kuehlmann and F.~Krohm, ``Equivalence checking using cuts and heaps,'' in
  \emph{Proceedings of the 34th ACM/IEEE Design Automation Conference},
  Anaheim, CA, June 1997, pp. 263--268.

\end{thebibliography}
\end{document}